\documentclass[%
 reprint, superscriptaddress,
 nofootinbib,
 amsmath, amssymb,
 aps, pra, longbibliography,
 floatfix, twocolumn%
]{revtex4-1}

\usepackage{graphicx}
\usepackage{dcolumn}
\usepackage[colorlinks=true,citecolor=blue,urlcolor=black]{hyperref}
\usepackage{subfigure}
\usepackage{color,soul}
\usepackage{braket}
\usepackage{mathtools, nccmath}
\usepackage{amsmath,amssymb}
\usepackage{enumerate}
\usepackage{multirow}
\usepackage{bm}

\newcommand{\Tr}{\operatorname{Tr}}
\let\nvec\vec
\def\vec#1{\nvec{\vphantom t\smash{#1}}}

\usepackage[dvipsnames]{xcolor}

\begin{document}
\title{Device-Independent Quantum Key Distribution with Random Postselection}

 \author{Feihu Xu}
 \affiliation{Hefei National Laboratory for Physical Sciences at Microscale and School of Physical Sciences, University of Science and Technology of China, Hefei 230026, China}
 \affiliation{CAS Center for Excellence in Quantum Information and Quantum Physics, University of Science and Technology of China, Shanghai 201315, China}
 \affiliation{Shanghai Research Center for Quantum Sciences, Shanghai 201315, China}
	
 \author{Yu-Zhe Zhang}
 \affiliation{Hefei National Laboratory for Physical Sciences at Microscale and School of Physical Sciences, University of Science and Technology of China, Hefei 230026, China}
 \affiliation{CAS Center for Excellence in Quantum Information and Quantum Physics, University of Science and Technology of China, Shanghai 201315, China}
 \affiliation{Shanghai Research Center for Quantum Sciences, Shanghai 201315, China}

 \author{Qiang Zhang}
 \affiliation{Hefei National Laboratory for Physical Sciences at Microscale and School of Physical Sciences, University of Science and Technology of China, Hefei 230026, China}
 \affiliation{CAS Center for Excellence in Quantum Information and Quantum Physics, University of Science and Technology of China, Shanghai 201315, China}
 \affiliation{Shanghai Research Center for Quantum Sciences, Shanghai 201315, China}

 \author{Jian-Wei Pan}
 \affiliation{Hefei National Laboratory for Physical Sciences at Microscale and School of Physical Sciences, University of Science and Technology of China, Hefei 230026, China}
 \affiliation{CAS Center for Excellence in Quantum Information and Quantum Physics, University of Science and Technology of China, Shanghai 201315, China}
 \affiliation{Shanghai Research Center for Quantum Sciences, Shanghai 201315, China}

\begin{abstract}
Device-independent quantum key distribution (QKD) can permit the superior security even with unknown devices. In practice, however, the realization of device-independent QKD is technically challenging because of its low noise tolerance. In photonic setup, due to the limited detection efficiency, a large amount of the data generates from no-detection events which contain little correlations but contribute high errors. Here we propose the device-independent QKD protocol with random post selection, where the secret keys are extracted only from the post-selected subset of outcomes. This could not open the detection loophole as long as the entropy of the post-selected subset is evaluated from the information of the entire set of data, including both detection and no-detection events. This post selection has the advantage to significantly reduce the error events, thus relaxing the threshold of required detection efficiency. In the model of collective attacks, our protocol can tolerate detector efficiency as low as $ 68.5\%$, which goes beyond standard security proofs. The results make a concrete step for the implementation of device-independent QKD in practice.
\end{abstract}

\maketitle

\emph{Introduction. ---}
Quantum key distribution (QKD)~\cite{bennett1984quantum,ekert1991quantum} allows two remote users, Alice and Bob, to share a secret key with information-theoretical security~\cite{xu2020secure,pirandola2020advances}.
The security of QKD normally relies on the assumption that the users's devices are trusted and well-characterized~\cite{lo1999unconditional,shor2000simple,renner2008security,scarani2009security}, but the imperfections in realistic devices may introduce potential loopholes~\cite{xu2010,Lars2010}. The measurement-device-independent QKD protocol~\cite{lo2012measurement} (see also an efficient version~\cite{lucamarini2018overcoming}) has been proposed to remove the side channels in measurement devices, where the state-preparation devices have to be trusted and calibrated~\cite{wei2020high}.

Device-independent QKD~\cite{mayers1998quantum,barrett2005no,acin2007device,pironio2009device,braunstein2012side}, as an entanglement-based protocol~\cite{ekert1991quantum}, relaxes conventional assumptions on the devices and allows the users to realize QKD with unkown and uncharaterized devices.
As long as some minimal assumptions are satisfied~\cite{pironio2009device,barrett2013memory,curty2017quantum}, the security of device-independent QKD can be guaranteed based solely on the violation of an Bell inequality. An intuitive understanding is that the violation of the Bell inequality certifies the presence of a quantum nonlocal state shared between Alice and Bob and consequently limits the information that can be obtained by the third party, Eve. Recently, theoretical efforts have advanced the developments of device-independent QKD for different scenarios~\cite{masanes2011secure,reichardt2013classical,vazirani2014fully,Arnon2018Practical}.

Unfortunately, device-independent QKD is challenging with current technology.
A practical realization of device-independent QKD typically requires that an Bell inequality is violated in a loophole-free fashion~\cite{hensen2015loophole,rosenfeld2017event}. A key problem in the photonic implementation is the restricted \emph{detection efficiency}, e.g., the emitted photons may not be detected due to the losses in the transmission or the detectors. Indeed, the detection efficiency determines the amount of violation of the Bell inequality and thus the possibility of secure key generation. Recently, researchers have closed the detection loopholes and demonstrated loophole-free Bell tests in photonic realizations~\cite{christensen2013detection,shalm2015strong,giustina2015significant,liu2018device} with detection efficiencies $\eta\sim  80\%$. Nevertheless, for the purpose of device-independent QKD, a much higher efficiency, e.g., $\eta > 90\%$, is required with the conventional security proofs~\cite{pironio2009device,masanes2011secure,reichardt2013classical,vazirani2014fully,Arnon2018Practical}, which is far beyond the current technologies. To lower the threshold efficiency, recent works have proposed different approaches, such as efficient post-processing~\cite{fourvalue}, two-way classical communication~\cite{tan2020advantage}, noisy preprocessing~\cite{ho2020noisy}, generalized Bell inequalities~\cite{woodhead2021device,sekatski2021device,gonzales2021device}, complete statistics via von Neumann entropy~\cite{brown2021computing,ComputeDI2021} and multiple key-generation basis~\cite{schwonnek2021device}.

Here we propose and prove the device-independent QKD protocol with random post selection. The basic idea is to extract secret keys only from a smaller string of outcomes corresponding to properly post-selected events, instead of from the entire data. The post-selection processing can effectively remove the no-detection events which contain little correlations but high errors~\cite{de2016randomness}. Note that this will \emph{not} evoke the detection loophole~\cite{christensen2013detection,shalm2015strong,giustina2015significant}, because the non-local test, i.e., the secret entropy, is evaluated from the entire data set that include all events. We prove the security against collective attacks, and show that the post selection can greatly reduce the information cost of error correction, thus facilitating the enhancement of loss tolerance. As an explicit result, we show that it is possible to reduce the threshold efficiency to $68.5\%$, which outperforms the standard security proof of $92.4\%$~\cite{pironio2009device} and the noisy preprocessing result of $82.6\%$~\cite{ho2020noisy,sekatski2021device} (see Table~\ref{Tab0}).

\begin{figure}
    \centering
    \includegraphics[width=1\linewidth,angle=0]{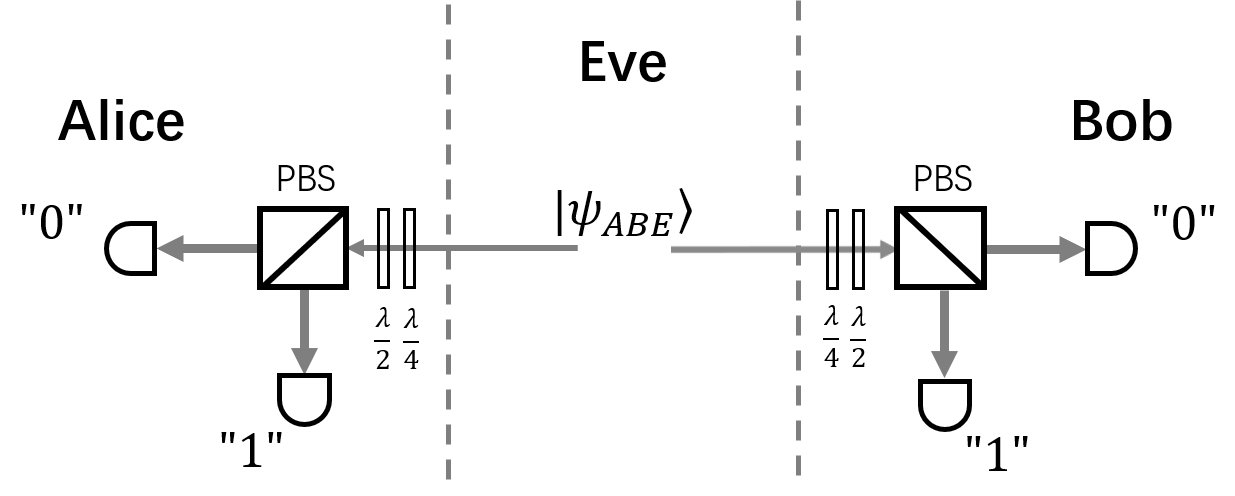}    \caption{\label{fig1} Photonic  realisation  of  the  device-independent QKD.  A quantum state created by a source that is potentially controlled by Eve is shared between Alice and Bob. Alice and Bob perform measurements using a polarising beam-splitter (PBS) and two detectors. A set of wave-plates $(\frac{\lambda}{4},\frac{\lambda}{2})$ allow them to choose the measurement setting. Each party effectively produces one of two possible outcomes $0$ and $1$.}
\end{figure}

\emph{Protocol.---}%
As shown in Fig.~\ref{fig1}, the device-independent QKD protocol we study is a modification of~\cite{pironio2009device}  using a photonic realisation. Two users, Alice and Bob, share a quantum channel consisting of a source which emits entangled photon pairs. After receiving the photons, Alice (or Bob) randomly chooses a measurement setting $x\in\{1,2\}$ (or $y\in\{1,2,3\}$) to measure the photon, and obtains the outcome $a$ and $b$.
For imperfect detectors, four events can be observed by Alice (or Bob): both detectors do not click, one detector clicks while the other one does not, and both detectors click. In the following, we label the event where only the first detector clicks as $``0"$ and the event where only the second detector clicks as $``1"$.
We assign the other two events, i.e. both detectors click or non-click, also as $``1"$.
In this manner, Alice and Bob each effectively produces one of two possible outcomes $a,b \in \{0,1\}$.

Let $P(a,b|x,y)$ denote the joint probability to obtain the output pair $(a,b)$ given the input pair $(x,y)$.
We randomly select a fraction of strings corresponding to the input pair $(\bar{x},\bar{y})=(1,3)$ as the ``key-generation round" to generate the secret key, while all the other strings, formulating the joint probabilities $P(a,b|x,y)$ for $x\in\{1,2\}$ and $y\in\{1,2,3\}$, are used as the ``test round" to characterize the nonlocal correlations. Our aim is to quantify the secret key conditional on the full set of nonlocal correlation $\{P(a,b|x,y)\}$, which is potentially more efficient than the evaluation based on the violation of a specific Bell inequality~\cite{nieto2014using,de2016randomness,RogerPRA2019}.

The core step in our protocol is the \emph{post-selection procedure}, where Alice and Bob randomly and independently post-select their own outcomes with respect to the key-generation rounds $(\bar{x},\bar{y})$. {Particularly, since the events associated with the two detectors both click or non-click at each side (labeled as $``1"$) contain little correlations, Alice and Bob will randomly and independently retain (or discard) bits $``1"$ with a probability $p$ (or $1-p$). Meanwhile, Alice and Bob will keep all the events associated with only the first detector clicks (labeled as $``0"$) which contain genuine quantum correlations in principle.} Note that for the ``test round", Alice and Bob keep all the outcomes without any post-selection such that the Bell test is done without detection loopholes. After the post-selection procedure, both Alice and Bob announce the discarded rounds using an authenticated public channel, and they only keep the bit pairs where both bits are retained for key generation. Note that since Eve can not access to the local randomness Alice and Bob use for post-selection, she does not know which round will be post-selected a priori~\cite{de2016randomness}.

The protocol is then proceeded with an error correction step that allows Bob to infer Alice's new (noisy) raw key.
The final secret key can be obtained after a privacy amplification step.

\emph{Key rate from the post-selected events. ---}
In the model of collective attacks~\cite{pironio2009device}, the devices behave in an independent and identically distributed (i.i.d.) manner.
Correspondingly, Eve can also extract information in an i.i.d. way by performing the individual measurement at each round. Let {$\mathcal{H}_A$, $\mathcal{H}_B$ and $\mathcal{H}_E$} be the Hilbert spaces of Alice’s device, Bob’s device and Eve’s device, respectively.
At the beginning of each round, a tripartite state $\rho_{ABE}$ is shared among Alice, Bob, and Eve.
After the measurements, the joint distribution of Alice's and Bob's outputs with respective to measurement settings can be described as
\begin{equation}
    P(a,b|x,y)=\Tr\left[(A_{a|x}\otimes B_{b|y} \otimes I)\rho_{ABE}\right].
\end{equation}
Here, $A_{a|x}$ and $B_{b|y}$ are the positive-operator-valued-measures associated with Alice's and Bob's measurements and outcomes.

Under post-selection, we use $\mathcal{V}_p$ represent the set of post-selected events, i.e., $\mathcal{V}_p=\{ab|ab=00,01,10,11\}$, and let $\omega_{00}=1$, $\omega_{01}=\omega_{10}=p$ and $\omega_{11}=p^{2}$. Given a bit pair, the probability that it can be kept is defined as $p_{\mathcal{V}_p}$: $p_{\mathcal{V}_p}=\sum_{ab\in\mathcal{V}}{\omega_{ab}P(a,b|\bar{x},\bar{y})}$.
After the post selection, the probability distributions $\hat{\bm{P}}$ of the post-selected events in $\mathcal{V}_p$ are given by,
\begin{equation}\label{0}
    \hat{P}(a,b|\bar{x},\bar{y},\mathcal{V}_p)=P(a,b|\bar{x},\bar{y})\cdot\omega_{ab}/p_{\mathcal{V}_p}.
\end{equation}
Conditioned on the post-selected events, the quantum side information can be represented by the state $\rho_{\hat{A}BE|\mathcal{V}_p}=\frac{1}{p_{{\mathcal{V}}_p}}\sum_{ab\in \mathcal{V}}\omega_{ab}|ab\rangle\langle ab|\otimes \rho_{ab}^{E}$, where $\rho_{ab}^{E}=\Tr_{AB}[(A_{a|\bar{x}}\otimes B_{b|\bar{y}}\otimes I)\rho_{ABE}]$.
By taking the partial trace over {$\mathcal{H}_B$}, we have
\begin{equation}
    \rho_{\hat{A}E|\mathcal{V}_p}=\frac{1}{p_{{\mathcal{V}}_p}}\sum_{a=0}^{1}|a\rangle\langle a|\otimes \left(\omega_{a,0}\rho_{a,0}^{E}+\omega_{a,1}\rho_{a,1}^{E}\right).
\end{equation}
Then, the conditional min-entropy of $\hat{A}$ given $E$ and the post-selected events $\mathcal{V}_p$ is determined by the guessing probability $G(\hat{A}_{\bar{x}}|E,\mathcal{V}_p)$, with which Eve can correctly guess the new raw key by measuring her side system $E$,
\begin{equation}
    H_{\text{min}}(\hat{A}_{\bar{x}}|E,\mathcal{V}_p)=-\log_{2}{G(\hat{A}_{\bar{x}}|E,\mathcal{V}_p)}.
\end{equation}
The  quantum  conditional  min-entropy quantifies  the strength of the correlation between Alice and Eve, and hence the secrecy of the key.

To evaluate the guessing probability, $G(\hat{A}_{\bar{x}}|E,\mathcal{V}_p)$, we examine the probability that Eve makes a correct guess $e=\hat{a}_{\bar{x}}$ of Alice's new post-selected raw keys, i.e.
\begin{equation} \label{Eq:guess}
  \frac{1}{p_{{\mathcal{V}}_p}}\Tr\left[\sum_{a}{M_{e=a}\cdot\left(\sum_{b}{\omega_{ab}\rho_{ab}^{E}}\right)}\right],
\end{equation}
where $\{M_{e}\}$ is Eve's measurement operators with $e\in\{0,1\}$ and it satisfies $M_0+M_1=I$. The maximal value of this guessing probability is determined by maximizing Eq.~\eqref{Eq:guess} over all quantum realizations $R=(|\psi\rangle,A_{\bar{x}},B_{\bar{y}},M)$, which are compatible with the given marginal $\{P(a,b|x,y)\}$~\cite{nieto2014using}.
Denoting $G_{\hat{a}|e}= \sum_{b}\omega_{ab} \langle \psi|A_{a|\bar{x}}\otimes B_{b|\bar{y}}\otimes M_{e}|\psi\rangle$, we thus have
\begin{align} \label{Eq:optimization}
    G(\hat{A}_{\bar{x}}|E,{\mathcal{V}}_p)=&\frac{1}{p_{{\mathcal{V}}_p}}\max_{R} \left({G_{0|0}+G_{1|1}}\right) \\ \nonumber
    &\text{s.t. } \langle \psi|A_{a|x}\otimes B_{b|y}\otimes I|\psi\rangle=P(a,b|x,y).
\end{align}

To solve the optimization program in Eq.~\eqref{Eq:optimization}, one can introduce a bipartite subnormalized quantum correlations
$P'_{e}(a,b|\bar{x},\bar{y})=\langle \psi|A_{a|\bar{x}}\otimes B_{b|\bar{y}}\otimes M_{e}|\psi\rangle$, which means the outcomes of Alice, Bob and Eve after the measurement are $a,b,e$ respectively. Hence, the problem in Eq.~\eqref{Eq:optimization} can be solved as semidefinite programs~\cite{NPA,NPA2}
\begin{align}
\label{2}
     G(\hat{A}_{\bar{x}}|E,{\mathcal{V}}_p)=&\frac{1}{p_{{\mathcal{V}}_p}}\max_{P'_{e}(a,b|\bar{x},\bar{y})}\sum_{ab\in \mathcal{V}}\omega_{ab}{P}'_{e=a}(a,b|\bar{x},\bar{y}) \\ \nonumber
     &\text{s.t. } \sum_{e\in\{0,1\}}P'_{e}(a,b|x,y)=P(a,b|x,y), \\ \nonumber
     & \quad \quad \quad P'_{e}(a,b|x,y) \in  \widetilde{Q},
\end{align}
where $ \widetilde{Q}$ denotes the set of unormalized bipartite quantum correlations~\footnote{Computations were performed with the NPA Hierarchy function in QET-LAB using the CVX package with solver Mosek.}.
Note that the constraints in Eq.~\eqref{2} involve all outcomes $(a,b)$. This respects the fact that detection loopholes are closed as no particular detection events are selected in the Bell test.

Combining the above analysis, the secret key rate $r$ in the asymptotic limit (and with optimal error correction) can be lower-bounded by the Devetak-Winter rate~\cite{DW2005},
\begin{equation}
\label{keyrate1}
    r\ge p_{\mathcal{V}_p}\left[H_{\text{min}}(\hat{A}_{\bar{x}}|E,\mathcal{V}_p)-H(\hat{A}_{\bar{x}}|B_{\bar{y}},\mathcal{V}_p)\right],
\end{equation}
where $H(A_{\bar{x}}|B_{\bar{y}},\mathcal{V}_p)$ is the cost of one-way error correction from Alice to Bob.

\emph{Simulation results. ---}
In the simulation, we focus on the threshold efficiency of the detection devices. We suppose that the devices are operated by using a pure and non-maximally entangled state which has the form $| \psi(\theta)_{AB} \rangle=\cos(\theta)|00\rangle +\sin(\theta)|11\rangle$, where $\theta \in [0,\pi/2]$. The corresponding density operator is denoted as $\rho_{\theta}=|\psi(\theta)\rangle \langle \psi(\theta)|$. For simplicity, we restrict measurements to be projective within the $x$--$z$ plane of the Bloch-sphere, i.e., measurements in the form of {
\begin{equation}
    \Pi(\phi)=\cos{(\phi)}\sigma_{z}+\sin{(\phi)}\sigma_{x},
\end{equation}
where $\phi \in [-\pi,\pi]$}. With the above notations, Alice and Bob's joint probability can be expressed as $p(a,b|x,y)=\Tr[\rho_{\theta} (A_{a|x}\otimes B_{b|y})]$, where $A_{a|x}$ and $B_{b|y}$ can be written as
\begin{align}
    A_{a|x}&=\delta_{a,0}\frac{1+\Pi(\phi_{x})}{2}\eta+\delta_{a,1}\left(\frac{1-\Pi(\phi_{x})}{2}\eta+(1-\eta)\right), \\ \nonumber
    B_{b|y}&=\delta_{b,0}\frac{1+\Pi(\phi_{y})}{2}\eta+\delta_{b,1}\left(\frac{1-\Pi(\phi_{y})}{2}\eta+(1-\eta)\right).
\end{align}
Here, $\eta \in [0,1]$ is the detection efficiency of the single-photon detectors.

\begin{figure}[htbp]
    \centering
    \includegraphics[width=0.95\linewidth,angle=0]{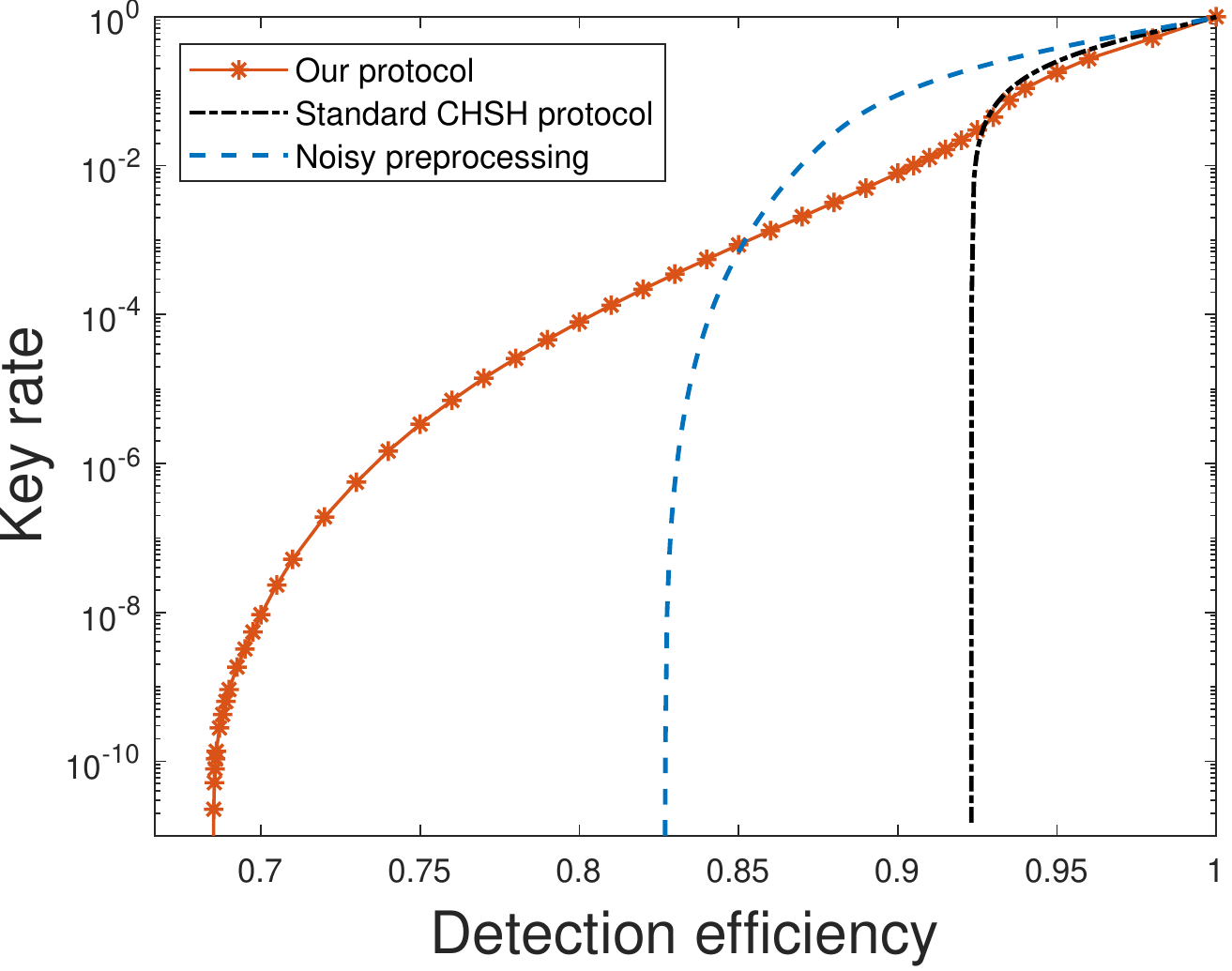}  \caption{\label{fig2} Asymptotic secret key rate as a function of detection efficiency. In the scenario of random post selection, the protocol can tolerate efficiency as low as $\eta \approx 68.5\%$ (red-solid curve). The black dot-dashed curve comes from Ref.~\cite{pironio2009device}, which shows positive key rate with threshold efficiency $\eta \approx 92.4\%$. The blue-dasher curve is obtained when using the protocol of noisy pre-processing~\cite{ho2020noisy,sekatski2021device}, and it shows a threshold efficiency of $\eta \approx 82.6\%$.}
\end{figure}

Firstly, we present an explicit numerical example to illustrate the practical advantage of our protocol. We choose a set of parameters which achieves a maximum key rate at detection efficiency $\eta=80\%$. The numerical result shows that the optimal entanglement parameter is $\theta=0.394$ and the measurements of Alice and Bob are with parameters $\{\phi_{x}\}=\{2.084,-2.853\}$ and $\{\phi_{y}\}=\{-2.272,2.926,-1.905\}$, respectively. Before the post selection, the cost of error correction is as large as $H(\hat{A}_{\bar{x}}|B_{\bar{y}},\mathcal{V}_p)=0.6501$. In contrast, after the post selection, the cost of error correction can be reduced to $H(\hat{A}_{\bar{x}}|B_{\bar{y}},\mathcal{V}_p)=0.03269$. More importantly, analog to noisy preprocessing~\cite{ho2020noisy}, the added randomness in post-selection can also limit Eve's guessing capabilities, i.e., increase the conditional min-entropy $H_{\text{min}}(\hat{A}_{\bar{x}}|E,\mathcal{V}_p)$. Here the optimal post-selection probability $p$ is 0.00527. In particular, without post selection, $H_{\text{min}}(\hat{A}_{\bar{x}}|E,\mathcal{V}_p)=0.03676$; but with post selection, $H_{\text{min}}(\hat{A}_{\bar{x}}|E,\mathcal{V}_p)=0.05914$. Finally, the net effect for the difference between $H_{\text{min}}(\hat{A}_{\bar{x}}|E,\mathcal{V}_p)$ and $H(\hat{A}_{\bar{x}}|B_{\bar{y}},\mathcal{V}_p)$, i.e., the secret key rate (see Eq.~\eqref{keyrate1}), is greatly enhanced, which achieves a positive key rate of $r\approx 7.9\times 10^{-5}$ bit per round. This example shows that our protocol can not only reduce the information cost of error correction, but also limit Eve's guessing capabilities.

Our main result is the red-solid curve as shown in Fig.~\ref{fig2}. By optimizing the entanglement parameter $\theta$ and the measurement settings, a positive key rate can be obtained even when the detection efficiency reaches at $68.5\%$. Here, we have considered that the error correction uses the three-valued outcome $b_{\bar{x}}$ \cite{fourvalue} (see Supplementary). The black dotted-dashed curve in Fig.~\ref{fig2} is the protocol in Ref.~\cite{pironio2009device}, where a minimum global detection efficiency of $\eta=92.4\%$ is needed. We also include the protocol of noisy preprocessing~\cite{ho2020noisy,sekatski2021device}, which requires the minimal detection efficiency around $\eta=82.6\%$ as shown by the blue-dashed curve. A comparison of threshold efficiency among different methods and proof techniques is shown in Table~\ref{Tab0}.

\begin{table}[ht!]
\caption{A comparison of threshold efficiency among different methods for device-independent QKD.} \label{Tab0}
\begin{tabular}{c @{\hspace{0.5cm}} c  }
  \hline \hline
   \textbf{Method} & \textbf{Threshold efficiency}\\
   \hline
   Standard analysis~\cite{pironio2009device} & 92.4\% \\
   Efficient post-processing~\cite{fourvalue} & 90.9\% \\
   Advantage distillation~\cite{tan2020advantage} & 89.1\% \\
   Iterated mean divergence~\cite{brown2021computing} & 84.5\% \\
   Noisy preprocessing~\cite{ho2020noisy} & 83.2\% \\
   Asymmetric inequality~\cite{sekatski2021device,woodhead2021device} & 82.6\% \\
   Quasi-relative entropy~\cite{ComputeDI2021} & 80.5\% \\
   This work &  68.5\% \\
\hline  \hline
\end{tabular}
\end{table}


\emph{Conclusion and Discussion. ---} Our security analysis uses the framework of min-entropy which might be extended via entropy accumulation theorem to finite-key analysis~\cite{Arnon2018Practical}.
{In Supplementary, we include the results after considering the device imperfections or noise, such as non-ideal source visibility. In comparison to noisy preprocessing~\cite{ho2020noisy}, the post-selection idea is slightly sensitive to noise. However, by using the analysis of von Neumann entropy~\cite{ComputeDI2021}, our approach can provide a stronger robustness~\cite{photonDI2021}. Furthermore, the post-selection idea can be combined with the recent proposals of complete statistics with numerical calculations~\cite{brown2021computing}, noisy preprocessing~\cite{ho2020noisy}, generalized Bell inequalities~\cite{woodhead2021device,sekatski2021device,gonzales2021device} and multiple key-generation basis~\cite{schwonnek2021device}, so as to tolerate higher loss. We leave those subjects to future works.}

Overall, we have proved the security of device-independent QKD with random post selections. Since the post-selected events have higher non-local correlations and lower errors, the protocol can achieve a significant reduction of the threshold detection efficiency. The high efficiency tolerance presents an important step towards the realization of device-independent QKD in practice. {Our proposal has been experimentally demonstrated using photonic implementations~\cite{photonDI2021}. Besides, we also notice two concurrent proof-of-concept device-independent QKD experiments based on trapped ions~\cite{ionDI2021} and trapped atoms~\cite{atomDI2021}.}

\section*{Acknowledgments}
We particularly thank Charles Lim for the motivation on the subject and the critical comments on the manuscript. We also thank Jean-Daniel Bancal, Peter Brown, Nicolas Sangouard, Ernest Tan, Wen-Zhao Liu, Yi-Zheng Zhen for helpful discussions. This work was supported by the National Natural Science Foundation of China (62031024), the National Key Research and Development (R\&D) Plan of China (2020YFA0309701), the Anhui Initiative in Quantum Information Technologies, Shanghai Municipal Science and Technology Major Project (2019SHZDZX01), Shanghai Academic/Technology Research Leader (21XD1403800) and the Chinese Academy of Sciences. F. Xu acknowledge the support from the Tencent Foundation.

F. Xu and Y.-Z. Zhang contribute equally. 

Contact: feihuxu@ustc.edu.cn (F.Xu).

%

%

\section*{Supplementary}

\subsection{The cost of one-way error correction with three-valued outcomes}
Considering the three possible outcomes corresponding to i) no click at all, ii) and iii) one click exactly in one of the two detectors, then $H(\hat{A}_{\bar{x}}|B_{\bar{y}},\mathcal{V}_p)$  can use the three-valued outcomes of $B_{\bar{y}}$ instead of its binarisation \cite{fourvalue}. In the following, we label that a click in the first detector as $``0"$, a click in the second detector as $``1"$ and no detection as outcome $``2"$. For the survived ``key generation" rounds, Alice and Bob will form \emph{the new raw keys} by assigning the no-detection events $``2"$ as a $``1"$. Since Bob knows which rounds are performed such binning, he can use this information to reduce the cost of one-way error correction.

We begin with the probability distributions $\hat{\bm{P}}$ of the post-selected events after Alice labels $``2"$ as $``1"$ at each survived rounds,
\begin{align}
\label{0}
    \hat{P}(00|\bar{x}\bar{y})=&P(00|\bar{x}\bar{y})/p_{{\mathcal{V}}_p}, \\ \nonumber
    \hat{P}(01|\bar{x}\bar{y})=&P(01|\bar{x}\bar{y})\cdot p/p_{{\mathcal{V}}_p}, \\ \nonumber
    \hat{P}(02|\bar{x}\bar{y})=&P(02|\bar{x}\bar{y})\cdot p/p_{{\mathcal{V}}_p}, \\ \nonumber
    \hat{P}(10|\bar{x}\bar{y})=&[P(10|\bar{x}\bar{y})+P(20|\bar{x}\bar{y})]\cdot p/p_{{\mathcal{V}}_p}, \\ \nonumber
    \hat{P}(11|\bar{x}\bar{y})=&[P(11|\bar{x}\bar{y})+P(21|\bar{x}\bar{y})]\cdot p^{2}/p_{{\mathcal{V}}_p}, \\ \nonumber
    \hat{P}(12|\bar{x}\bar{y})=&[P(12|\bar{x}\bar{y})+P(22|\bar{x}\bar{y})]\cdot p^{2}/p_{{\mathcal{V}}_p}.
\end{align}
Here, $p_{{\mathcal{V}}_p}=\sum_{ab\in\mathcal{V}}{\omega_{ab}P(ab|\bar{x}\bar{y})}$ with $\omega_{00}=1$, $\omega_{01,02,10,20}=p$ and $\omega_{11,12,21,22}=p^2$. For simplicity, we are going to write $\hat{P}(ab|\bar{x}\bar{y})$ as $\hat{P}_{ab}$. The cost of one-way error correction $H(\hat{A}_{\bar{x}}|B_{\bar{y}},\mathcal{V}_p)$ can be calculated as
\begin{align}
    H(\hat{A}_{\bar{x}}|B_{\bar{y}},\mathcal{V}_p)=& h(\hat{P}_{00})+h(\hat{P}_{01})+h(\hat{P}_{02}) \\ \nonumber +&h(\hat{P}_{10}) +h(\hat{P}_{11})+h(\hat{P}_{12}) \\ \nonumber 
    -&h(\hat{P}_{00}+\hat{P}_{10})-h(\hat{P}_{01} \\ \nonumber
    +&\hat{P}_{11})-h(\hat{P}_{02}+\hat{P}_{12})
\end{align}
where $h(x)$ is defined as $h(x)=-x\cdot\log_{2}(x)$.

\subsection{{Simulation for the scenario with device imperfections}}
{
In this section, we derive a noisy model for the photonic realization to consider the device imperfections, and perform the simulation for the DI-QKD protocol with random postselection. Our model will consider the imperfections in the measurement devices such as dark counts, and the imperfection in the entanglement source such as the non-ideal fidelity and multiple photon pairs.}

{
For a real entanglement source, we use the visibility to quantify the prepared state, i.e.,
\begin{equation}
 \rho_{AB}=V\times |\psi(\theta)\rangle \langle \psi(\theta)| +(1-V)\times \frac{I}{4},
\end{equation}
where $V$ denotes the visibility along a certain measurement direction and $| \psi(\theta) \rangle=\cos(\theta)|00\rangle +\sin(\theta)|11\rangle$ is the distributed non-maximally entangled state. So far, a visibility value as high as $99.2\%$ has been demonstrated in experiment~\cite{photonDI2021}. We also consider that the photon source may emit multiple pairs of photons which follows a Poisson distribution. For simplicity, we will restrict to projective measurements within the $x$--$z$ plane of the Bloch-sphere. As for the key generation rounds, Bob will record four kinds of outcomes to perform the error correction, i.e., i) and ii) one click exactly in one of the two detectors, iii) both the two detectors click and iv) no click at all.}

{
Note that we have proven the security of random post-selection by bounding the min-entropy $H_{\text{min}}(\hat{A}_{\bar{x}}|E,\mathcal{V}_p)$ between Alice and Eve. However, this approach typically leads to a sub-optimal bound on the von Neumann entropy and it is not noise-robust. To consider the tolerance to device imperfections, we adopt the method in Ref.~\cite{ComputeDI2021} to obtain a tighter bound for the von Neumann entropy $H(\hat{A}_{\bar{x}}|E,\mathcal{V}_p)$. Further details can be seen in Ref.~\cite{photonDI2021}.}

\begin{figure}[htbp]
    \centering
    \includegraphics[width=0.5\linewidth,angle=0]{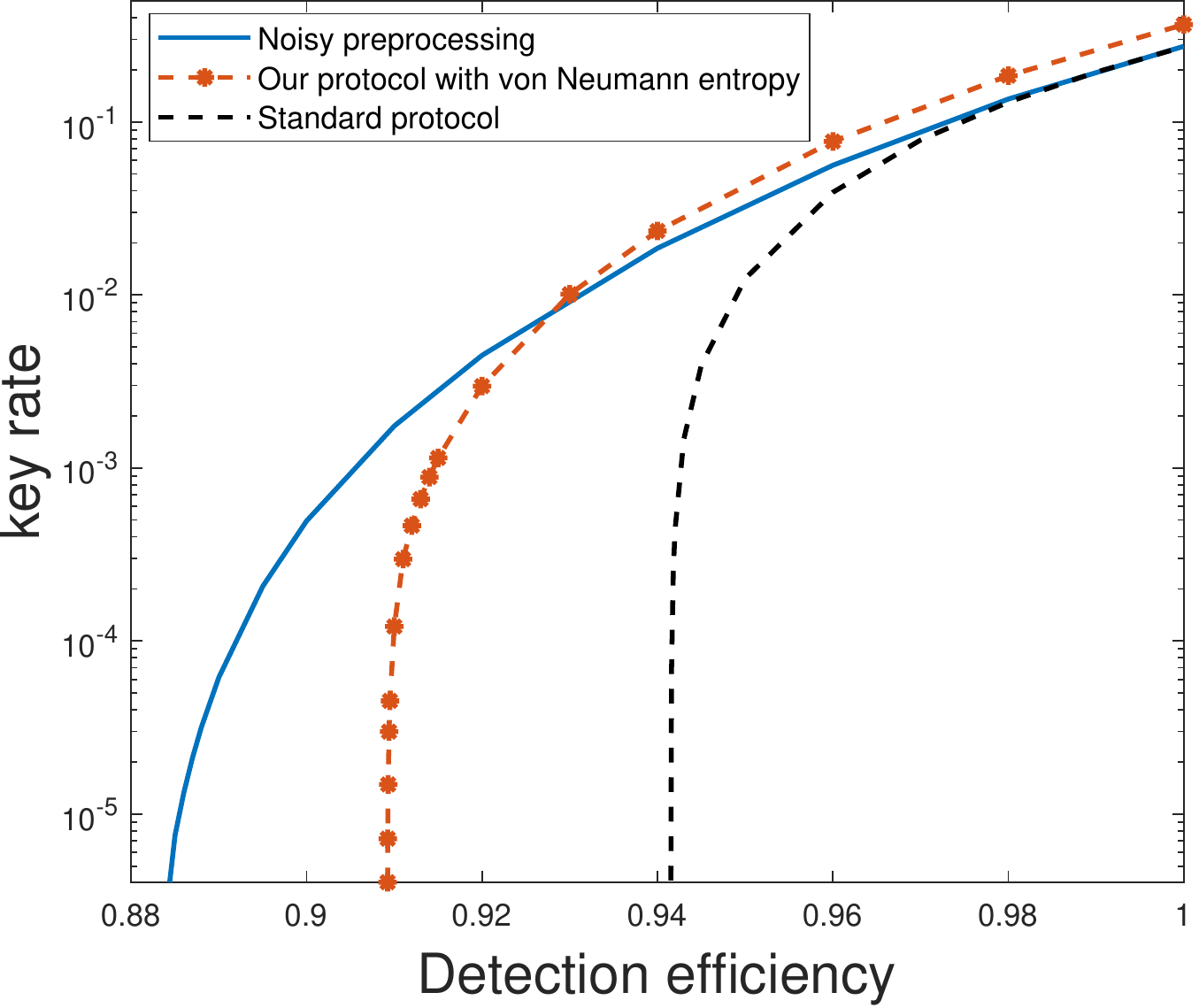}  \caption{{\label{fig3} Asymptotic secret key rate as a function of detection efficiency. We consider multiple photon pairs as input, the imperfect visibility of about 99.2\%, and the dark count probability of a single detector of $10^{-6}$. In this case, our protocol can tolerate efficiency as low as $\eta \approx 90.9\%$ (red-dasher curve). The black dashed curve comes from Ref.~\cite{pironio2009device}, which shows positive key rate with threshold efficiency $\eta \approx 94.1\%$. The blue-solid curve is obtained when using the protocol of noisy pre-processing~\cite{ho2020noisy,sekatski2021device}, and it shows a threshold efficiency of $\eta \approx 88.3\%$.}}
\end{figure}

{
We consider multiple photon pairs as input, the imperfect visibility of about 99.2\%, and the dark count probability of a single detector of $10^{-6}$. After performing numerical simulations to optimize the parameters, our main result is the red-dashed curve as shown in Fig.~\ref{fig3}, i.e., a positive key rate can be obtained when the detection efficiency reaches at $90.9\%$. The black dashed curve in Fig.~\ref{fig3} is the standard CHSH protocol~\cite{pironio2009device}, where a minimum global detection efficiency of $94.1\%$ is needed. For the protocol of noisy preprocessing~\cite{ho2020noisy,sekatski2021device}, it has a better performance to tolerate the noise, which requires the minimal detection efficiency around $88.3\%$ as shown by the blue-solid curve.
}

\end{document}